\documentstyle[11pt]{article}
\begin{document}

\begin{centering}
{\Large Results from recent searches for muonium-- antimuonium conversion and 
future perspectives}\\
\end{centering}
\vspace*{5mm}

\begin{centering}
{K.~Jungmann $^*$\\
\vspace*{3mm}
Physikalisches Institut der Universit\"at
Heidelberg\\
Philosophenweg 12\\
 D-69120 Heidelberg, Germany\\}
\end{centering}

\vspace*{10mm}
A positive muon ($\mu^+$) and an electron ($e^-$) form the the hydrogen-like
muonium atom ($M$=$\mu^+ e^-$). Since it consists of two leptonic particles
which are according to present
knowledge point-like, accurate calculations of its level energies
can be performed in the framework of standard theory. In particular,
the theoretical description of this system is possible  almost exclusively
by Quantum Electrodynamics (QED) which has been verified in various
very high precision experiments [1].
Due to the close confinement of the bound state
the muonium atom
renders in addition the possibility for sensitive searches for electron--muon
interactions not included in the standard model. Such exotic
processes could affect the energy of the quantum states at a level
of the same order of magnitude as the precision achievable
in modern spectroscopic experiments.
Oscillations between muonium and antimuonium ($\overline{M}$=$\mu^- e^+$)
would violate separate
additive lepton number conservation which is a purely empirical law.\\

Whereas in the hadronic sector quark mixings
are well known ($K^0$--$\overline{K^0}$ oscillations
are a familiar example),
they have not been observed for leptons, yet.
$M$--$\overline{M}$-oscillations are not provided in the standard model,
however, they are allowed in many extensions to
it which try to explain further some of its features like, e.g.
parity violation or the spectrum of elementary particle masses.
In many of these speculative models the
coupling constant ${\rm G_{M\overline{M}}}$ in an effective
four fermion interaction
could be as high as the present upper limit
established in an experiment at LAMPF in Los Alamos, USA, of ${\rm
G_{M\overline{M}}} \leq 0.16 {\rm G_F}$~ (90\%C.L.) [2] or the result
of ${\rm G_{M\overline{M}}} \leq 0.14 {\rm G_F}$~ (90\%C.L.)
[3]
reported recently from
an experiment at the Phasotron in Dubna, Russia,
where
${\rm G_{F}}$ is the Fermi coupling constant.\\
 
At the Paul Scherrer Institut (PSI) in Villigen, Switzerland, a new sensitive
experiment has been set up and successfully concluded a  series of
measurements [4,5] in which decays of $\overline{M}$-atoms were searched after
producing $M$ atoms from a beam of positive muons by electron capture in a
SiO$_2$ powder target. 
The signature requested for the observation of
$M$--$\overline{M}$-oscillations
comprises the coincident
detection of both constituents of the antiatom in its decay.
A $\mu^-$--decay
($\mu^-\rightarrow e^-+\overline{\nu}_e+\nu_{\mu}$)
releases an energetic electron with an energy spectrum ranging up to 53 MeV
and can be identified in a magnetic
spectrometer consisting of five cylindrical
proportional wire chambers in coaxial geometry operated at 0.1~T field.
The positron in the atomic shell
of the antiatom is left behind after the decay with an average
kinetic energy of $R_{\mu}$=13.5~eV. This particle
is accelerated to 7~keV and transported
in a magnetic guiding field onto a position sensitive microchannel plate
detector (MCP). The 511~keV radiation from the annihilation
of positrons in the MCP
can be detected in a segmented pure CsI crystal detector.
The muonium atoms are formed by stopping
a beam of positive muons
in a $ {\rm SiO_2}$ powder.
A fraction of about 7\% of the incoming muons
is converted into muonium atoms which diffuse to the surface and leave the
powder with thermal energies into the surrounding vacuum [6,7].
It is essential for the experiment to have the atoms in vacuum,
since the $M$--$\overline{M}$-oscillation is strongly suppressed
in gases or condensed matter due to the removal of symmetry
between the atom and the antiatom.
The muonium yield is monitored regularly
by reversing the electric acceleration field and all magnetic fields
approximately every five hours
and by observing energetic positrons and electrons respectively.
These measurements also serve for calibrating detcetor subsystems.
The targets are replaced about every third day to account for an
observed time dependent decrease of the muonium production efficiency,
because of changes in the target surface structure and contaminations.
Attention has been paid to keep the apparatus as much as possible symmetric
for detecting muonium and antimuonium decays.\\
 
From the data recorded in a first
stage of the experiment which contained no event fulfilling the requested
signature
the formerly
established upper bound [2,3] on the probability of the process 
could be improved
by almost
two orders of magnitude.
Assuming a ${\rm (V\pm A)\times(V\pm A)}$
interaction type a new limit on the coupling constant of
${\rm G_{M\overline{M}}} \leq  1.8 \cdot 10^{-2}~G _{\rm F}$ (90\% C.L.)
could be established [4].
 
In a second and advanced stage of the experiment data have been  
collected for 4 month.
Various experimental improvements had been implemented
among which were increased detection efficiencies
for both charged particles originating from the
atoms decay, reduced background in the detector components,
higher muon beam fluxes by using the $\pi$E5 beam area of PSI
and extended running time. For the positron detector an increase
in efficiency by a factor of 4 could be achieved by employing
a suited secondary emitting foil in front of the microchannel plate detector
[8].
The analysis is completed and is presently being cross checked. One event
was found within 3 standard deviations of all relevant parameters in 
good agreement with an expected background of 1.7(2) events from accidental 
coincidences.
The upper limit for the coupling constant in a ${\rm (V\pm A)\times(V\pm A)}$
type interaction could be set to
${\rm G_{M\overline{M}}} \leq  3 \cdot 10^{-3}~G _{\rm F}$ (90\% C.L.).\\

This result rules out a $Z_8$ model with
radiative mass generation by heavy lepton seed [10].
It  allows further to set
a lower limit on the mass of dileptonic gauge bosons in GUT theories
of 2.6~TeV/c$^2$ * g$_{3l}$ which is significantly 
above values extracted from high energy Bhabha
scattering experiments [11].
In the framework of R-parity violating supersymmetric models [12,13] the bound
on the coupling parameters could be lowered by a factor of 15 
to $\mid \lambda_{132}\lambda_{231} \mid \leq 3* 10^{-4}$ with assumed
superpartner masses of order 100 GeV/c$^2$.
The achieved level of sensitivity allows to slightly narrow the 
interval of allowed 
heavy muon neutrino masses
in minimal left-right symmetric
models [14] ( which have predicted a lower bound on
${\rm G_{M\overline{M}}}$) to $\approx$ 40~keV/c$^2$ up to the present
experimental bound at 170~keV/c$^2$.\\
 
From an experiment using a significantly brighter pulsed muon source 
but a similar event signature one
could expect an increase in sensitivity almost proportional to the
increase in the muon flux. It will be essential to take advantage 
of the time dependence of a possible muonium to antimuonium conversion
process [15]. The oscillation process is expected to start increasing 
with a square dependence in time wheras major background is expected to decay
exponentially in time. The signal could be searched for after a
significant part of the background has decayed. For such an experiment a 
time structure 
of muon pulses of up to about 4~$\mu$sec length and more than 10to 15~$\mu$sec
separation would be ideal. It could be realized from a muon source derived 
from the high energy hadron machine at JHF.\\

This work is supported by the German Bundesminister f\"ur Bildung und Forschung
(BMBF), the Russian Federation for Fundamental Research (RFFR) and the Swiss
Nationalfond.\\
 
\begin{flushleft} \small
 $^*$ This paper is based on the work of the 
 muonium-antimuonium collaboration at PSI which at present consists of:\\
V.~Baranov,
V.~Karpu\-chin,
I.~Kisel,
A.~Korenchen\-ko,
S.~Korenchen\-ko,
N.~Kravchuk,
N.~Kuchinsky,
A.~Moiseenko,
{\it JINR Dubna};
A.~Gro\ss{}mann, 
K.~Jungmann,
J.~Merkel,V.~Meyer,
G.~zu~Putlitz,
I.~Reinhard,
P.V.~Schmidt,
K.~Tr\"ager,
L.~Willmann,
A.~Gro\ss{}mann, 
K.~Jungmann,
J.~Merkel,V.~Meyer,
G.~zu~Putlitz,
I.~Reinhard,
P.V.~Schmidt,
K.~Tr\"ager,
L.~Willmann,
{\it University of Heidelberg};
R.~Abela,
D.~Renker,
H.K.~Walter,
{\it Paul Sherrer Institute};
J.~Bagaturia,
D.~Mzavia,
T.~Sakhelash\-vil\-li,
{it Tbilisi University};
V.W.~Hughes,
{it Yale University};
R.~Engfer,
H.P.~Wirtz,
{\it University of Z\"urich}\\
\vspace*{5mm}

{\normalsize \bf References}
\\

\vspace*{5mm}
\begin{itemize}
\item[1] V.W.~Hughes and G.~zu~Putlitz, in: "Quantum
Electrodynamics", p. 822ff, T.~Kinoshita (ed.), World
Scientific, Singapore
(1990)
\item[2]  B.E.~Matthias et al., Phys.Rev.Lett. {\bf66}, 2716 (1991)
\item[3] V.A. Gordeev et al., JETP Lett. {\bf 59}, 589 (1994)
\item[4] R. Abela et al., Phys.Rev.Lett.{\bf 77}, 1950 (1996); 
sea also:
K.~Jungmann et al., in:
"Nuclear and Particle Physics
with Meson Beams in the 1GeV/c Region",
S.~Sugimoto and 
O.~Hashimoto (eds.)
Universal Academy Press, Tokyo, p. 137 (1995);
K.~Jungmann,
Atomic Physics 
14, D.J. Wineland, C.E. Wieman and S.J. Smith (eds.),
AIP Press,
New York, p. 102
\item[5] W.~Bertl et al., in: "Weak and Electromagnetic Interactions",
H.~Eijiri (ed.), World Scientific, p. 117 (1995)
\item[6] K.~Woodle et al., Z.Phys. D{\bf9}, 59 (1988);
\item[7] A.C.~Janissen et al., Phys.Rev. A{\bf42}, 161 (1990)
 \item[8] P.V. Schmidt et al., Nucl.Instr.and Meth. {\bf a 376}, 139 (1996)
 \item[9] V. Meyer et al., Proceedings of the 6th Conference on Intersections of 
Particle and Nuclear Physics, Big Sky, Montana, USA (1997)
 \item[10] G.G. Wong and W.S. Hou, Phys.Lett. B{\bf357}, 145 (1995)
 \item[11] K. Horikawa and K. Sasaki, Phys.Rev. D{\bf 53}, 560 (1996)
 \item[12] R. Mohapatra, Z. Phys. {\bf C56}, S117 (1992) 
 \item[13] A. Halprin and A. Masiero, Phys.Rev. {\bf D 48}, 2987 (1993)
 \item[14]  P.~Herczeg and R.N.~Mohapatra, Phys.Rev.Lett. {\bf69}, 2475 (1992)
 \item[15] L.~Willmann and K.~Jungmann, in: "Atomic Physics Methods 
in Modern Research", K. Jungmann, J. Kowalski, I. Reinhard and F. Tr\"ager 
(eds.), Springer, p. 43 (1997)
\end{itemize}
\end{flushleft}
 
\end{document}